\documentclass[nohyperref]{article}

\usepackage{microtype}
\usepackage{graphicx}
\usepackage{subfigure}
\usepackage{booktabs} % for professional tables

\usepackage{hyperref}

\newcommand{\bb}{\textbf}
\usepackage[accepted]{icml2022}

\usepackage{amsmath}
\usepackage{amssymb}
\usepackage{mathtools}
\usepackage{amsthm}
\usepackage{natbib}
\usepackage{xspace}

\usepackage[capitalize,noabbrev]{cleveref}

\theoremstyle{plain}

\theoremstyle{definition}

\theoremstyle{remark}

\usepackage[textsize=tiny]{todonotes}
\newcommand{\mytitle}{Accelerating Distributed Deep Learning using Lossless Homomorphic Compression}

\newcommand{\ie}[0]{\textit{i.e.,}\xspace}
\newcommand{\eg}[0]{\textit{e.g.,}\xspace}

\def\Snospace~{\S{}}

\icmltitlerunning{\mytitle{}}

\begin{document}
\pagestyle{plain}
% \fancyfoot[C]{\thepage}
\bibliographystyle{plainnat}
% \bibliography{mybib}
% \bibliographystyle{unsrt} 



\twocolumn[
\icmltitle{\mytitle{}}
\icmlsetsymbol{equal}{*}

\begin{icmlauthorlist}
\icmlauthor{Haoyu Li}{ut}
\icmlauthor{Yuchen Xu}{pku}
\icmlauthor{Jiayi Chen}{ut}
\icmlauthor{Rohit Dwivedula}{ut}
\icmlauthor{Wenfei Wu}{pku}
\icmlauthor{Keqiang He}{airbnb}
\icmlauthor{Aditya Akella}{ut}
\icmlauthor{Daehyeok Kim}{ut}
\end{icmlauthorlist}

\icmlaffiliation{ut}{UT Austin.}
\icmlaffiliation{pku}{Peking University.}
\icmlaffiliation{airbnb}{Airbnb}

\icmlcorrespondingauthor{Haoyu Li}{lhy@utexas.edu}
\icmlcorrespondingauthor{Aditya Akella}{akella@cs.utexas.edu}
\icmlcorrespondingauthor{Daeheyeok Kim}{daehyeok@utexas.edu}

\icmlkeywords{Machine Learning, ICML}]

\vskip 0.3in



\printAffiliationsAndNotice{}  % leave blank if no need to mention equal contribution

\begin{abstract}
As deep neural networks (DNNs) grow in complexity and size, the resultant increase in communication overhead during distributed training has become a significant bottleneck, challenging the scalability of distributed training systems. Existing solutions, while aiming to mitigate this bottleneck through worker-level compression and in-network aggregation, fall short due to their inability to efficiently reconcile the trade-offs between compression effectiveness and computational overhead, hindering overall performance and scalability. In this paper, we introduce a novel compression algorithm that effectively merges worker-level compression with in-network aggregation. Our solution is both homomorphic, allowing for efficient in-network aggregation without CPU/GPU processing, and lossless, ensuring no compromise on training accuracy. Theoretically optimal in compression and computational efficiency, our approach is empirically validated across diverse DNN models such as NCF, LSTM, VGG19, and BERT-base, showing up to a 6.33$\times$ improvement in aggregation throughput and a 3.74$\times$ increase in per-iteration training speed.

%\haoyu{Deep neural networks (DNNs) have been rapidly growing in size due to the increase in computing power and availability of large datasets. This growth necessitates distributed training on extensive clusters, where communication overhead has emerged as a bottleneck. The bottleneck occurs as distributed workers frequently need to aggregate local gradient updates. Current approaches, including worker-level compression and in-network aggregation, attempt to address the bottleneck from different angles, but have respective drawbacks. An appealing idea is to combine these approaches to harness the benefits of both, but it is also challenging primarily due to hardware limitations.  In this paper, we propose a novel worker-level compression algorithm compatible with in-network aggregation, thereby resolving the conflict. Specifically, our algorithm has two main properties. The first, homomorphic, allows for aggregation after compression without CPU/GPU involvement, increasing the aggregation throughput. The second, lossless, ensures to reduce traffic to be communicated without loss of information, eliminating the impact of training accuracy. Theoretically, we prove that our algorithm achieves asymptotically optimal compression ratios and computational complexity. Empirically, we demonstrate a significant performance improvement of distributed training across various tasks, including NCF, LSTM, VGG19 and BERT-base, accelerating up to 6.33$\times$ in aggregation throughput and 3.74$\times$ in per iteration training speed.}
\end{abstract}

\section{Introduction}\label{sec_introduction}
%\haoyu{TODO: Para 4 needs to add drawbacks of existing methods; Para 5 needs motivation of compact property; Para 6 needs key insight of compact property; Para 7 needs rephrase to correspond to the homomorphic, lossless and compact properties respectively.}
% In recent years, deep neural networks (DNNs) have been trained with ever-increasing parameter counts due to the
% rapid growth of computing power and large datasets. For example, BERT-base and BERT-large had 110 million and
% 340 million parameters~\cite{devlin2018bert}, respectively. Today, GPT-3 has over 175 billion parameters~\cite{brown2020language}. This vast quantity of parameters enables models to capture a broader range of high-level, non-linear features;
% pivotal in advancing various fields, including image processing~\cite{he2016deep}, natural language processing~\cite{yin2017comparative}, and recommendation systems~\cite{gupta2020architectural} among others.

Deep neural networks (DNNs) have been rapidly growing in size due to the increase in computing power and availability of large datasets. Models like BERT-base and BERT-large have 110 million and 340 million parameters respectively~\cite{devlin2018bert}, while GPT-3 boasts an impressive 175 billion parameters~\cite{brown2020language}. Larger models can capture a wider range of high-level features, leading to advances in fields such as image and natural language processing and recommendation systems~\cite{he2016deep, yin2017comparative,gupta2020architectural}.

% Because loading all the training data and the model in one single machine seems impractical \rohit{don't think we need this first phrase motivating why DDP}, distributed training has emerged as an inevitable trend, where a dominant and simplest way is data parallelism. 

% Distributed training (\eg data parallelism)\jane{Are we discussing the other types of distributed training, why or why not, should clearly state that} has emerged as the default\jane{can't say that} answer to dealing with the ever-growing scale of neural models. In data parallelism, data is split across multiple workers \cite{ben2019demystifying}, which compute local gradients using stochastic gradient descent (SGD) methods \cite{bottou2018optimization}, then execute a global aggregation\jane{is synchronization the right word? aggregation?}, typically summation, of local gradients with a centralized Parameter Server \cite{li2014scaling,jiang2020unified, luo2018parameter} or decentralized AllReduce \cite{chan2007collective,patarasuk2009bandwidth, barnett1994interprocessor} aggregation architecture. 

Distributed training using data parallelism~\cite{ben2019demystifying} is a standard approach for large DNN models. Local gradients are computed through stochastic gradient descent (SGD)~\cite{bottou2018optimization} by multiple workers, and then aggregated (summed up) globally using either a centralized parameter server~\cite{li2014scaling,luo2018parameter,jiang2020unified} or a decentralized AllReduce approach~\cite{barnett1994interprocessor,chan2007collective,patarasuk2009bandwidth}.

However, it is challenging to aggregate gradients due to the large volume of traffic generated, particularly when network bandwidth is limited. This can create a bottleneck, especially when dealing with large models. For instance, it takes almost half the time for aggregation when training BERT-base with eight workers \cite{devlin2018bert}. Furthermore, GPU computation throughput is advancing faster than network bandwidth \cite{sun2019optimizing}, which means that the existing imbalance between computation and communication will continue to grow \cite{sevilla2022compute}.

Prior work tackles the communication bottleneck in two ways. First, worker-level compression techniques~\cite{wang2023communication} aim to reduce the amount of data being communicated, thereby enabling gradient aggregation at a throughput that exceeds the network bandwidth. However, their compressed data formats often require additional CPU/GPU processing during aggregation, which might offset the speed gains achieved from reduced traffic ~\cite{li2023thc}.
Second, in-network aggregation approaches delegate the aggregation process to network infrastructure components such as programmable hardware switches~\cite{sapio2021scaling, lao2021atp} and specialized middleboxes~\cite{gebara2021network}. They take advantage of high packet processing speeds and bandwidth capabilities of switches and middleboxes, effectively bypassing CPU/GPU involvement and enhancing training throughput~\cite{feng2023network}. Nevertheless, this training throughput is strictly bounded by the physical network bandwidth, and the constraint makes it increasingly difficult to address the growing imbalance between computation and communication.

%the reliance on customized hardware \dk{prog switches are not customized hardware. also, need to describe why using HW make it difficult} makes it extremely difficult to implement optimizations beyond the simple aggregation of gradient vectors. \dk{is it true for the middlebox work?}
%\dk{need to describe drawbacks of each approach to motivate their combination in the next para.}

% There are two mainstream approaches trying to eliminate the bottleneck. The first approach is host-level compression \cite{wang2023communication}, like gradient sparsification and quantization \cite{hoefler2021sparsity}. For example, O$k$-Top$k$ \cite{li2022near} shows that the worker can select only 1\% parameters for aggregation without significant loss of accuracy; OmniReduce \cite{fei2021efficient} leverages the commonality of sparse gradients in DNNs by transmitting only non-zero parameters for aggregation, potentially reducing traffic by up to 99.3\%. The second approach is in-network aggregation, which offloads the aggregation process to network layer hardwares, like programmable switches \cite{sapio2021scaling, lao2021atp} and specialized middleboxes \cite{gebara2021network}. Compared to primitive aggregation methods, in-network aggregation capitalizes on the high packet processing speeds and bandwidth available in switches,  significantly enhancing the training throughput \cite{feng2023network}.
%\cite{hoefler2021sparsity} gives a detailed overview of sparsification in deep learning.
%\dk{avoid using citations as nouns.}

It is tantalizing to combine and harness the benefits of both 
%\dk{what's the benefits of each approach; aren't both reducing the traffic volume?} 
to optimize communication; however, these two approaches encounter inherent conflicts primarily stemming from hardware limitations~\cite{li2023thc}, making their seamless integration challenging. Compression algorithms usually do not support for aggregation after compression, referred to as ``homomorphic compression'' \cite{li2023thc,jang2017homomorphic}\footnote{The term ``homomorphic'' means the compressed aggregated data is equivalent to the aggregated compressed data.}. Even for those that do, such as Sketched-SGD~\cite{ivkin2019communication}, the use of probabilistic data structures inevitably introduces accuracy loss. This loss may necessitate additional iterations for the model to converge, potentially undermining the overall training performance~\cite{li2023thc}. From the perspective of in-network aggregation, challenges arise when attempting to directly aggregate data compressed in formats like coordinate list, commonly used in compression algorithms.

% However, the two approaches face inherent conflicts, primarily due to hardware limitations \cite{li2023thc}. Specifically, in-network switches find it challenging to directly aggregate data compressed in formats like COO (coordinate list), which are commonly used in compression algorithms. Most compression methods, like OmniReduce \cite{fei2021efficient} and Top-$k$A \cite{renggli2019sparcml}, presume the presence of a host machine to pre-process the data before aggregating them. Although there are some compression algorithms supporting aggregation after compression, as referred to ``homomorphic compression''\footnote{The term ``homomorphic'' means the compressed aggregated data is equivalent to the aggregated compressed data.}, like Sketched-SGD \cite{ivkin2019communication}, their probabilistic data structures inevitably incur accuracy loss. This, in turn, necessitates more iterations for the model to converge, potentially undermining overall training performance \cite{li2023thc}. In summary, the pursuit of efficient compression, in-network computing, and the lossless property appears to be a challenging task.

%\jane{too many details; edited, ptal} 
In this paper, we propose a homomorphic and lossless compression algorithm that can work with in-network gradient aggregation without incurring accuracy loss.
%\dk{you should describe key insights / high-level ideas first. Below seems too details. } 
%
At a high level, our algorithm consists of two phases to aggregate gradients: compression and recovery. In the compression phase, each worker encodes the original gradient, denoted as \bb{\textit{X}}, into a homomorphic compressed form, denoted as $\bb{\textit{S}}(\bb{\textit{X}})$, and transmits $\bb{\textit{S}}(\bb{\textit{X}})$ to the existing aggregation API. In the recovery phase, each worker receives the aggregated compressed data, \ie{} $\bb{\textit{S}}(\sum\bb{\textit{X}})$, from the aggregation API and accurately recovers $\sum\bb{\textit{X}}$ from it.
The novelty of our algorithm hinges on two key techniques. First, we use two homomorphic data structures to construct $\bb{\textit{S}}(\sum\bb{\textit{X}})$ while preserving all information. Second, we utilize a practical algorithm from the field of random graphs for the accurate recovery of the aggregated gradient.
To elaborate, our initial step involves compressing the original gradient \bb{\textit{X}} into a homomorphic dimensionality-reducing structure known as Count Sketch~\cite{charikar2002finding}.
Following this, we let $\bb{\textit{S}}(\sum\bb{\textit{X}})$ includes not only the Count Sketch but also an additional homomorphic indexing structure. This indexing structure is pivotal in identifying non-zero parameters, thereby mitigating potential estimation errors on zero values. More importantly, we observe that after discarding the zero parameters, the size of non-zero parameters of DNN gradients can be comparable to the size of the Count Sketch. This suggests that Count Sketch operates as a de facto invertible linear transformation instead of a tool for dimensionality reduction, preserving information without any loss. 
Finally, we discover that an existing algorithm in random graphs, known as parallel peeling~\cite{jiang2017parallel} can be adeptly applied to recover the aggregated data, and thus achieving the lossless property.

Our algorithm theoretically boasts two compelling features beyond its homomorphic and lossless properties. First, our algorithm achieves a high compression ratio, defined as the size of the original data divided by the size of the compressed data. In fact, for any data type (\eg 4-bit integer) and any level of sparsity (\eg when 99\% parameters are zero) of the parameters, it guarantees asymptotically optimal use of bottleneck network bandwidth.  %and is GPU friendly.
%
%\dk{GPU friendliness comes from nowhere. is this your design requirement? if so, mention it earlier} 
%
Second, our algorithm achieves a high computation throughput with linear (asymptotically optimal) computational complexity for compression and decompression. The processes also exhibit good spatial locality, further enhancing their computational efficiency.
%\rohit{o(1)?}
%
%In cases where this threshold is not met, the algorithm still provides an unbiased estimation of the parameters as a backup.

We empirically demonstrate a remarkable enhancement in the performance of distributed training across various tasks, including Neural Collaborative Filtering (NCF) \cite{he2017neural}, Long Short-Term Memory networks (LSTM) \cite{hochreiter1997long}, VGG19 \cite{simonyan2014very}, and BERT-base \cite{devlin2018bert}. We integrate our algorithm into the most popular NVIDIA Collective Communications Library (NCCL) AllReduce framework and the ATP \cite{lao2021atp} in-network aggregation framework in PyTorch \cite{paszke2019pytorch}, finding that compared to the baseline training performance using standard NCCL and ATP, our algorithm achieves up to 6.33$\times$ speedup in terms of aggregation throughput and 3.74$\times$ speedup in terms of per iteration training. We have open-sourced our implementation.\footnote{\url{https://github.com/lihy0529/lossless_homomorphic_compression}.}

\section{Related Work}\label{related_work}
%\textcolor{red}{//THIS SECTION HAS NOT BEEN FINISHED}

%\haoyu{Flow: (a) PS and AllReduce archs. (b) Compression and INA techs.}

%As the data and model size increases, Parameter Server and AllReduce are two popular aggregation archtectures in data parallelism. 

We overview the two mainstream approaches to mitigate the communication bottleneck: compression and in-network aggregation. %For a comprehensive discussion, we direct to two detailed surveys on them \cite{wang2023communication,feng2023network} for reference.

Compression is a method of reducing data volume to decrease communication traffic using two techniques. The first, sparsification, involves filtering out parameters of lesser magnitude before transmission for aggregation. For example, selecting only the top 1\% of parameters in terms of absolute value for aggregation does not significantly impact accuracy~\cite{li2022near}. 
Various studies in sparsification, such as Top$k$A~\cite{wang2020fft}, Top$k$DSA~\cite{renggli2019sparcml}, and gTop$k$~\cite{shi2019distributed}, focus on the efficient selection of either local or global Top$k$ parameters. However, these works generally lack support for aggregation after compression, referred to as the homomorphic property~\cite{jang2017homomorphic,li2023thc}, due to their reliance on specific parameter storage formats like the coordinate list. This necessitates decompressing the gradients from all workers, aggregating them, and then compressing again before redistribution, significantly increasing computational overhead and traffic latency \cite{li2023thc}. Even in works like Sketched-SGD~\cite{ivkin2019communication}, which compress gradients into a homomorphic data structure called Count Sketch, they suffer from accuracy loss due to their probabilistic nature, leading to more training iterations and reduced model accuracy. 
Our compression algorithm, in contrast, seeks to maintain accuracy while preserving the homomorphic property. 
The second key technique in compression is gradient quantization, which involves mapping the value of each parameter to a lower precision range. For example, TernGrad~\cite{wen2017terngrad} shows that mapping each parameter's value into a two-bit range $\in\{-1,0,1\}$ can still preserve training convergence. Similar to sparsification, many quantization studies vary mainly in their bit-width approach, such as FP16~\cite{micikevicius2017mixed}, 8-Bit~\cite{dettmers20158}, and SignSGD~\cite{seide20141}. However, as these works are orthogonal to our algorithm, they will not be the focus of our discussion.

In-network aggregation approaches offload the aggregation process to network infrastructure components such as programmable hardware switches and specialized middleboxes. They leverage the high packet processing speed of switches and middleboxes to aggregate up to terabits of parameters per second. This approach mitigates the communication bottleneck by minimizing the traffic that would originally transverse the entire network. Examples of systems employing this method include SwitchML~\cite{sapio2021scaling} and ATP~\cite{lao2021atp}. 
However, in-network aggregation has a significant drawback in the form of hardware restrictions, as switches and middleboxes have a limited number of pipeline stages available for processing fixed-function operations~\cite{li2023thc}. This makes it challenging to directly aggregate data in formats other than simple gradient vectors.
Even in works like Libra~\cite{pan2022libra} which support aggregation after compression, they inevitably incur additional accuracy loss because they cannot proceed all parameters within limited memory on switches. Our compression algorithm, in contrast, seeks to support in-network aggregation on compressed gradients while keeping the lossless property.

\section{Algorithm Design}\label{sec_algorithm}

We propose a lossless homomorphic compression algorithm, simultaneously reducing the volume of data to be communicated without information loss, as well as avoiding CPU/GPU involvements during aggregation, thus harnessing the benefits of both compression and in-network aggregation.

At a high level, our algorithm consists of two phases to aggregate gradients (Algorithm \ref{algorithm}): compression and recovery. In the compression phase, each worker encodes the original gradient, denoted as \bb{\textit{X}}, into a homomorphic compressed form, denoted as $\bb{\textit{S}}(\bb{\textit{X}})$, and transmits $\bb{\textit{S}}(\bb{\textit{X}})$ to the existing aggregation API such as NCCL AllReduce and ATP in-network aggregation~\cite{lao2021atp}. 
In the recovery phase, each worker receives the aggregated compressed data, \ie $\bb{\textit{S}}(\sum\bb{\textit{X}})$, from the aggregation API and accurately recovers $\sum\bb{\textit{X}}$ from it. 
$\bb{\textit{S}}(\bb{\textit{X}})$ comprises two fundamental as well as homomorphic components: a data structure \bb{\textit{Y}} called Count Sketch, satisfying \bb{\textit{Y}}($\sum$\bb{\textit{X}})$=\sum$\bb{\textit{Y}}(\bb{\textit{X}}), and a compact indexing structure \bb{\textit{B}}, satisfying \bb{\textit{B}}($\sum$\bb{\textit{X}})$=\vee$\bb{\textit{B}}(\bb{\textit{X}}), where $\vee$ represents aggregating gradients with bitwise-OR operation. %Later in this section, we will formally define the terms Count Sketch, the index, and the \bb{peel($\cdot$)} function in the pseudo code.

\begin{algorithm}
\renewcommand\baselinestretch{0.9}\selectfont
  \caption{Lossless Homomorphic Compression}\label{algorithm}
   \bb{Phase I: Compression}\\
   \bb{Input:} gradient {\bb{\textit{X}}};\\
   \text{\quad Build Count Sketch } \bb{\textit{Y}} of {\bb{\textit{X}}};\texttt{ //\S\ref{homomorphic}}\\
   \text{\quad Build index } \bb{\textit{B}} of {\bb{\textit{X}}};\texttt{ //\S\ref{peeling_theory}, \S\ref {bloom_filter}}\\
   \bb{\quad Output:} \bb{aggregate} (\bb{\textit{S}}(\bb{\textit{X}}) $:=[\bb{\textit{Y}}, \bb{\textit{B}}]$); \texttt{ //API}\\

    \texttt{// The API makes } \bb{\textit{S}}(\bb{\textit{X}}) $\leftarrow$ \bb{\textit{S}}($\sum$\bb{\textit{X}})\\ 
    \texttt{// by making} \bb{\textit{Y}} $\leftarrow$ $\sum$\bb{\textit{Y}} \texttt{ and } \bb{\textit{B}} $\leftarrow$ $\vee$\bb{\textit{B}}.\\

   \bb{Phase II: Recovery}\\
   \bb{Input:} $[\bb{\textit{Y}}, \bb{\textit{B}}]:=$ \bb{\textit{S}}(\bb{\textit{X}});\texttt{ //API}\\
   \text{\bb{\quad\textit{X}}}$\leftarrow$\bb{peel} (\bb{\textit{Y}}, \bb{\textit{B}}); \texttt{ //\S\ref{peeling_theory}}\\
   \text{\quad}Estimate the not recovered parameters of \bb{\textit{X}} from \bb{\textit{Y}};\\
   \bb{Output:} \bb{\textit{X}}; \texttt{ //Aggregated gradient}
\end{algorithm}

This section is structured as follows: In \autoref{homomorphic}, we leverage sketching theory \cite{li2022stingy} to achieve the homomorphic property. We employ Count Sketch \cite{charikar2002finding}, which is pivotal in providing an unbiased estimation of parameters. In \autoref{peeling_theory}, we leverage peeling theory \cite{jiang2017parallel} to achieve the lossless property. 
We incorporate an additional bitmap index alongside the Count Sketch, accurately reconstructing the parameters from their compressed state. In \autoref{bloom_filter}, we leverage membership theory \cite{li2023chainedfilter} to achieve a high compression ratio, defined as the size of the original data divided by the size of the compressed data. We theoretically propose using a probabilistic data structure called Bloom Filter \cite{bloom1970space} as the additional index, making the compression ratio asymptotically optimal.\footnote{This is only theoretically necessary when the sparsity is extremely high.} In \autoref{locality}, we make the locality optimization to achieve high computation throughput.  We substantially reduce the number of memory access by processing consecutive parameters in batches.

\subsection{Count Sketch: Homomorphic Compression}\label{homomorphic}

Count Sketch~\cite{charikar2002finding} is an effective technique for unbiased dimensionality reduction. We start from here because it meets the homomorphic property, \ie satisfies $\vec{Y}(\sum\vec{X})=\sum\vec{Y}(\vec{X})$. Specifically, it compresses the original $\vec{X}$ into an array $\vec{Y}$ in the following manner: Each index $i$ of $\vec{X}$ is hashed to three signals $g_1(i), g_2(i), g_3(i) \in \{-1,1\}$, and three different indexes $h_1(i), h_2(i), h_3(i)$ in $\vec{Y}$. The Count Sketch process involves incrementing the $h_j(i)$-th ($j=1,2,3$) parameter in $\vec{Y}$ by $(g_j(i)\cdot x_i)$, which results in either $X_i$ or $-X_i$, depending on the signal $g_j(i)$. To estimate aggregated gradient $\vec{X}$, the worker examines the mapped value $Y_{h_j(i)}$ for each index $i$ in $\vec{X}$.  Ideally, if there is a one-to-one mapping between $X_i$ and $Y_{h_j(i)}$, the value of $X_i$ can be deduced as $g_j(i) \cdot Y_{h_j(i)}$. However, hash collision, where $Y_{h_j(i)}$ accumulates values from multiple counters, can occur. To mitigate this, Count Sketch reports the median of the values  $g_j(i)\cdot Y_{h_j(i)}$ as its estimation.  Including the $g_j(i)$ factor ensures that all potential collisions distribute symmetrically with a zero mean, thereby maintaining the unbiased nature of the estimation. %An example of Count Sketch is depicted in \bb{Figure \ref{figure_count_sketch}}.

Now we can formally describe a Count Sketch-based compression algorithm that is homomorphic, though not yet lossless. The algorithm involves two phases to aggregate gradients. In the compression phase, each worker encodes the original gradient $\vec{X}$ into a Count Sketch $\vec{Y}$, and transmits $\vec{Y}$ to the existing aggregation API. In the recovery phase, each worker receives the aggregated Count Sketch $\sum\vec{Y}$, and estimate the parameters of $\sum\vec{X}$ from it.

%In compression phase, each worker transmits the Count Sketch $\vec{Y}$ as the compressed data for aggregation. In recovery phase, each worker receives an aggregated Count Sketch $\vec{Y}\leftarrow\sum\vec{Y}$. Note that each worker uses identical hash mappings, so that corresponding values across different workers are correctly aligned for summation (aka ``homomorphic''). 

% \begin{figure}[h!tbp]
%   \centering
%    \includegraphics[width=1\linewidth]{icml2022/count_sketch.pdf}
% \caption{An example of count sketch. In each bipartite graph, the left part is the sparse data, while the right part is the sketch array. To insert the second parameter of the sparse data, we update the first ($g_1(2)=1$), second ($g_2(2)=1$), and fifth ($g_5(2)=-1$) counter of count sketch. We insert the fourth and fifth parameters in the same manner. To query an parameter, such as the second parameter of the sparse data, we report $median \{1\times3,1\times (-1), -1\times(-3)\}.$}
% \label{figure_count_sketch}
% \end{figure}

\subsection{Parallel Peeling: Lossless Recovery}\label{peeling_theory}

To achieve the lossless property, we need to leverage an additional but very common feature of DNN gradient, called sparsity.
This feature refers to the fact that most of the parameters are zero~\cite{fei2021efficient}. 
By recording the locations of these zero parameters with a small index, we can compress only the non-zero parameters in the Count Sketch, making lossless compression possible. %otherwise, compressing while preserving all information contravenes information theory.

We attain the lossless property by having each worker transmit an additional bitmap $\vec{B}$ as the index, allocating one bit per parameter - for each bit,   true indicates non-zero, while false indicates zero. It's easy to find that the bitmap is also homomorphic in terms of bitwise-OR ($\vee$) operation, \ie $\vec{B}(\sum\vec{X})=\vee\vec{B}(\vec{X})$, as the aggregation of multiple non-zero parameters results in a non-zero value.\footnote{Admitting that there's a tiny probability that the result is zero, it doesn't affect the correctness of our algorithm.} Therefore, incorporating the compressed data with an additional bitmap, \ie $\vec{S}(\vec{X})=[\vec{Y},\vec{B}]$, does not violate the homomorphic property, as long as we employ arithmetic summation to aggregate the Count Sketch and employ bitwise-OR to aggregate the bitmap, respectively. %\aditya{what is zip()?}

%Subsequently, the network consolidates all these bitmaps using a bitwise-OR operation ($\vee$).
% At this time, we let every worker node sends an additional bitmap of sparse data to indicate their non-zero parameters. The network aggregates (merges) all bitmaps with bitwise OR ($\vee$) and thus the target node receives the bitmap of the aggregated data. Next we show how the target node recovers the aggregated data with the help of the bitmap.
%
The bitmap serves dual purposes. An obvious one is to directly indicate zero parameters and exactly report them. The second purpose is more intriguing: It enables a one-to-one mapping between the non-zero parameters and the Count Sketch. %This process, namely peeling \cite{jiang2017parallel}, is shown in Figure \ref{figure_peeling_theory} and explained below.

\begin{figure}
  \centering
   \includegraphics[width=1\linewidth]{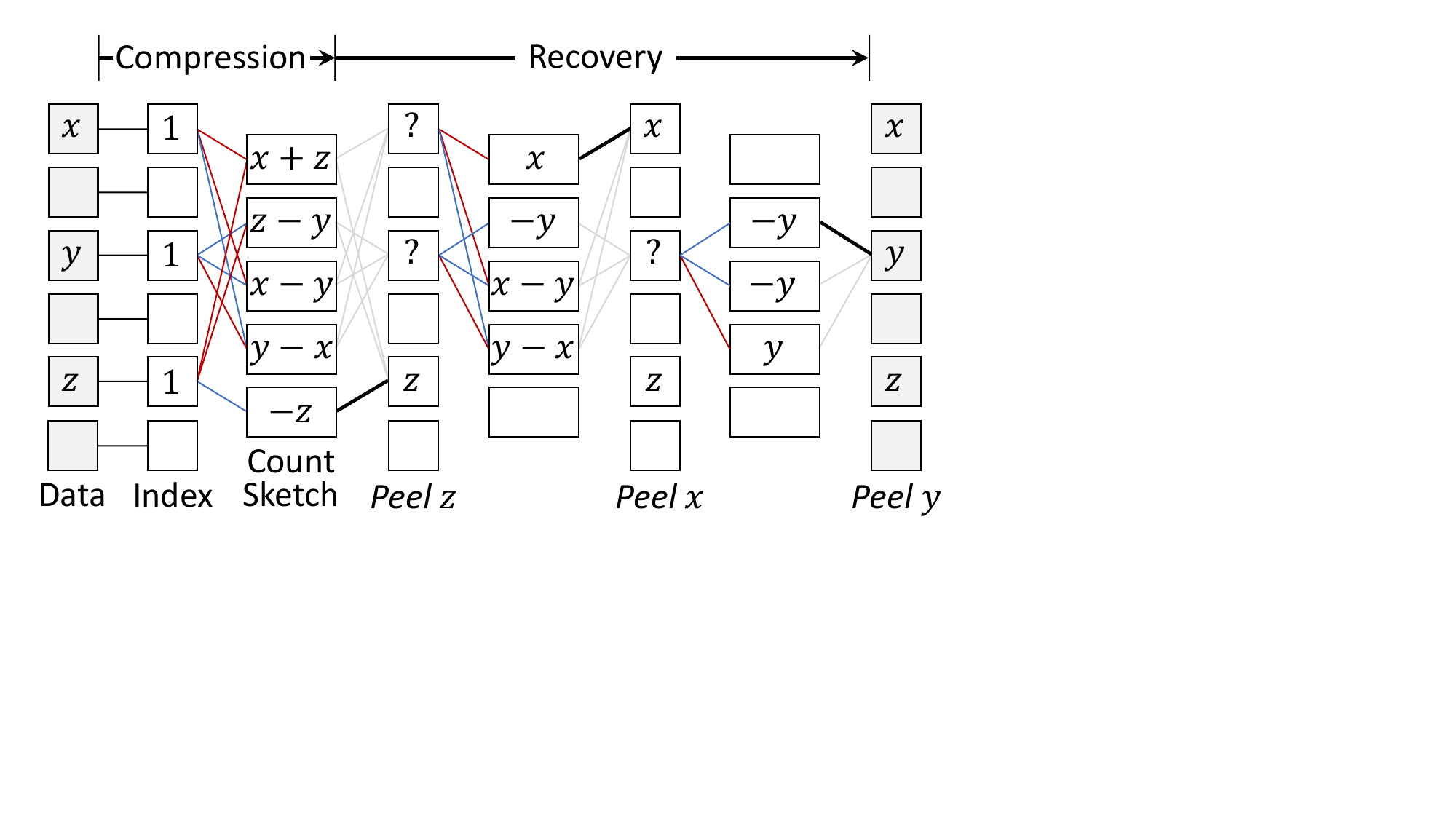}
   \vspace{-0.2in}
\caption{An example of our lossless homomorphic compression algorithm. In this example, three non-zero parameters - $x, y,$ and $z$ -  identified by the index, are hashed and added to the Count Sketch. Ignoring the aggregation process and the parallel computation for the moment, to recover the data from the index and the Count Sketch, we first peel away $z$, since its value uniquely hashes to the bottom counter of the Count Sketch. Following this removal, $x$ and $y$ become eligible for peeling. This sequence allows for the lossless recovery of the entire gradient.}
\label{example}
\end{figure}

Since we know the indexes of all non-zero parameters, we can figure out which indexes of $\vec{Y}$ are mapped by only one non-zero parameter in $\vec{X}$. For these parameters, we accurately recover their value $X_i$. Subsequently, we remove them from the bitmap by resetting the corresponding bit to zero. At the same time, we remove them from the Count Sketch by deducting $Y_{h_j(i)}$ by $g_j(i)\cdot X_i$. This process is akin to peeling away $X_i$ from its insertion. After the two removal steps, there may emerge new peelable parameters, allowing us to continue the peeling process until $\vec{X}$  is completely reconstructed or there are no more parameters that can be peeled.\footnote{For the not peelable parameter, we use the Count Sketch to give an unbiased estimation.} 
We show an example of our lossless homomorphic compression algorithm in \autoref{example}.
%For the remaining not recovered parameters, should they exist as indicated by the bitmap, we can provide an unbiased estimation in the same manner as the Count Sketch, or transmit them again.

Theoretically, the peeling process is equivalent to finding two cores of a bipartite graph. Two closely related works include IBLT \cite{goodrich2011invertible} and Counter Braids \cite{lu2008counter}. The distinction between these works and ours lies in the application focus: IBLT concentrates on compact key-value pair storage, whereas our approach supports homomorphic updates of values. Conversely, Counter Braids is tailored for network measurement with positive integers, while our method facilitates gradient aggregation for various data types. 

According to the parallel peeling theory \cite{jiang2017parallel}, we can fully recover the aggregated data with a high probability of $1-O(1/n)$, thus achieving the lossless property. This holds true provided that the size of the Count Sketch is at least $\gamma n$ where $\gamma=1.23$ and $n$ represents the number of non-zero parameters. Furthermore, it is proved that the parallel peeling process typically requires only $\log\log n +O(1)$ iterations to complete, and this figure can be easily optimized to $O(1)$ iterations in our algorithm by splitting the Count Sketch into multiple blocks of fixed size, which is good news for the computation throughput.

\subsection{Bloom Filter: High Compression Ratio}\label{bloom_filter}

Though the design in~\autoref{peeling_theory} is already lossless, the bitmap presents a theoretical challenge: its size becomes significant, particularly for data types with small bit-width, when dealing with extremely high sparsity. To mitigate this issue, we substitute the bitmap with a probabilistic indexing structure called Bloom Filter \cite{bloom1970space}, thereby rendering the compression ratio asymptotically optimal.

Before delving into the details, let us reconsider the concept of lossless compression and establish a theoretical lower bound for the required storage space, denoted as $S_{\min}$. Suppose $N$ is the total number of parameters of bit-width $C$ and $n$ is the number of non-zero parameters with values ranging from $[1..2^C-1]$. For complete recovery of all parameters, a compressed data structure must include at least two types of information: an index of size $S_1$ bits identifying non-zero parameters and a table of $S_2$ bits recording their corresponding values. In scenarios where $S_1=0$, the index does not distinguish zero and non-zero values, necessitating the table to record all values as though the data were dense. Conversely, when $S_1=N$, the index precisely identifies all zero and non-zero values, and the table only needs to record non-zero values. We therefore assume a balance between these extremes, with the index incorrectly identifying a certain portion, say $\epsilon$, of zero parameters as non-zero. Consequently, the table records $\epsilon(N-n)$ redundant values. We denote $\epsilon$ as the false positive rate, and the aggregated data contains $\lambda n:= N - n$ zero parameters. According to the chain rule theory of ChainedFilter \cite{li2023chainedfilter}, the overall space cost $S$ satisfies %\rohit{something is wrong with the math; define C}
\begin{equation*}
    \begin{split}
            \begin{cases}
            S   &\geqslant S_1 +S_2;\\
            S_1  &\geqslant nf(\epsilon,\lambda) + o(n);\\
            S_2  &\geqslant \log \left[\binom{Cn}{Cn(1+\epsilon\lambda)}\cdot (2^C-1)^n\right]\\&=Cnf(0,\epsilon\lambda)+n\log(2^C-1)+o(n);\\
            f(\epsilon,\lambda)&:=f(0,\lambda)-f(0,\epsilon\lambda);\\
            f(0,x) &:= (x+1)H(\frac{1}{x+1}), \forall x>0;\\
            H(x) &:= -x\log x-(1-x)\log (1-x), \forall x\in (0,1);
            \end{cases}\\
            \Rightarrow S\geqslant (C-1)nf(0,\epsilon\lambda)+nf(0,\lambda)+n\log(2^C-1)\quad&\\
            +o(n)\geqslant nf(0,\lambda)+n\log(2^C-1):=S_{\min}\quad\quad&
    \end{split}
\end{equation*}
%
% Specially, when $C=1$, the right hand side degenerates to the theoretical lower bound of the exact membership problem.

% \begin{figure}[h!tbp]
%   \centering
%    \includegraphics[width=1\linewidth]{icml2022/bloom_filter.pdf}
% \caption{An example of Bloom filter. In each bipartite graph, the left part is the sparse data, while the right part is the bit array of Bloom filter. To insert any parameter of the sparse data, we set the mapped bits to one (shown as black cells). To query an parameter, we report the bitwise-and ($\wedge$) result of all mapped bits.}
% \label{figure_bloom_filter}
% \end{figure}
%
Next, we illustrate how to asymptotically approach this bound $S_{\min}$ using a probabilistic indexing structure called Bloom Filter \cite{bloom1970space}. A Bloom Filter is essentially a probabilistic bit array, with a size determined by the formula $\frac{n}{\ln2}\log 1/\epsilon$ bits.\footnote{In this context, $n$ refers to the number of non-zero parameters and $\epsilon$ is the false positive rate mentioned in the previous paragraph. For simplicity, we don't round formulas to the nearest integers.} 
In the compression phase, each worker node constructs its Bloom Filter by hashing every non-zero parameter to $\log 1/\epsilon$ bits within the Bloom Filter and setting them to one. In the recovery phase, workers can approximate whether a parameter is non-zero by verifying that all the Bloom Filter's corresponding bits are set to one. An indispensable nature of Bloom Filter to achieve the lossless property is that it never recognizes non-zero values as zero. This is because if any hashed bit of a parameter is zero, the parameter is zero.  Otherwise, all the parameter bits would have been set to one when inserted. Another important property is that the Bloom Filter is homomorphic in terms of bitwise-OR ($\vee$) operation, \ie $\vec{B}(\sum\vec{X})=\vee\vec{B}(\vec{X})$, so the aggregation API can merge the Bloom Filters like handling bitmaps.  %\textbf{Figure \ref{figure_bloom_filter}} demonstrates this Bloom filter construction process with an example.
%A Bloom Filter is a probabilistic bit array of $\frac{n}{\ln2}\log 1/\epsilon$ bits to identify non-zero parameters with false positive rate $\epsilon$. In the compression phase, each worker randomly maps every non-zero parameter to $\log 1/\epsilon$ bits of the Bloom Filter and set them to one. Because aggregating Bloom Filters with bitwise-OR operations doesn't incur false negatives, the aggregation system can merge Bloom Filters just like treating bitmaps. In recovery phase, By checking whether the mapped bits are all one, the each worker can approximately identify whether an parameter is non-zero by checking whether the mapped bits are all one $\epsilon$. \bb{Figure \ref{figure_bloom_filter}} uses an example to show the construction process of Bloom filter. 
%

\begin{figure}[]
  \centering
   \includegraphics[width=1\linewidth]{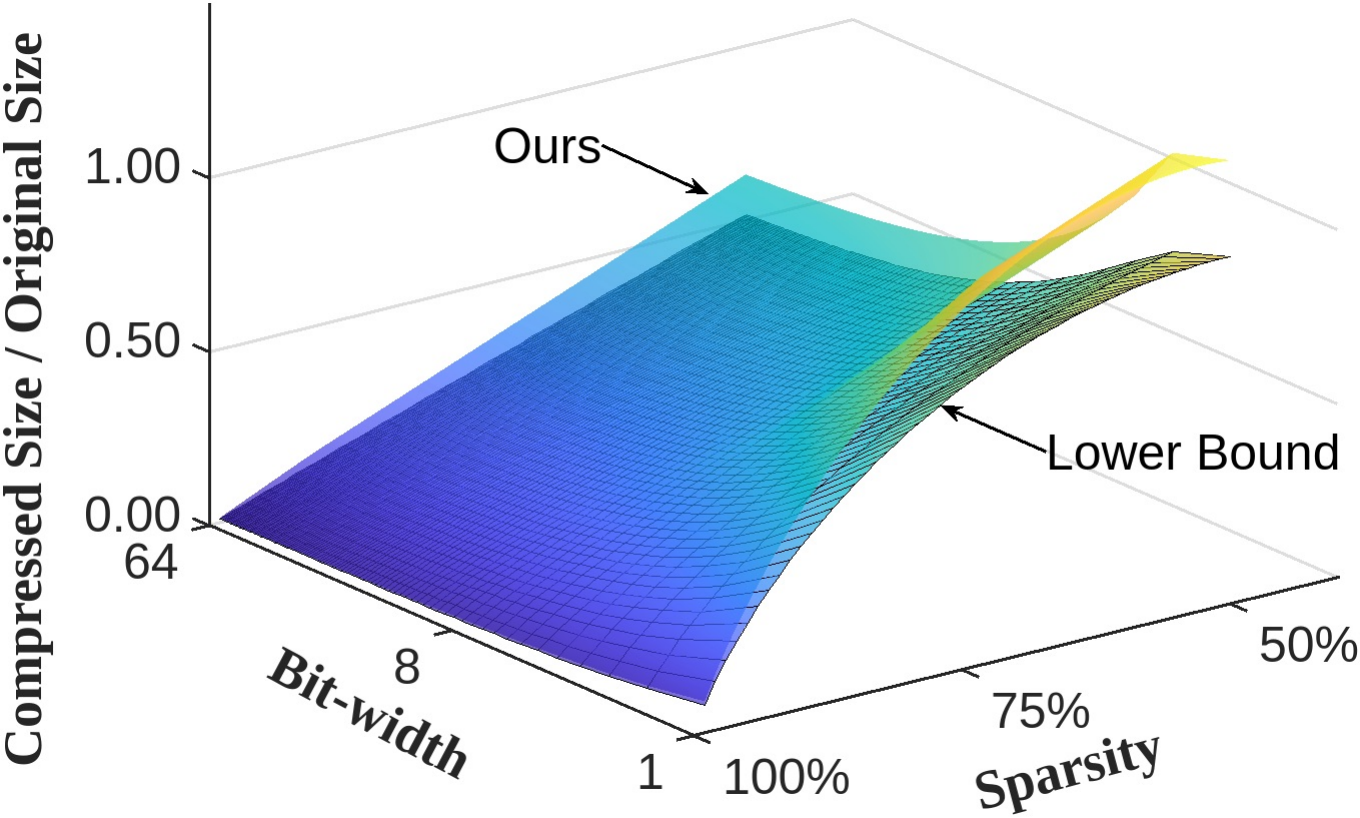}
   \vspace{-0.2in}
\caption{Theoretical compressed data size.}
\label{pic:theory}
\end{figure}

By setting $\epsilon$ to $(\ln^22\gamma C\lambda)^{-1}\leqslant 1$, where $\gamma$ is $1.23$ mentioned in \autoref{peeling_theory}, we can figure out the overall space cost of the Count Sketch and the Bloom Filter $S$ satisfies
\begin{equation*}
\begin{split}
\begin{cases}
S &= S_1 +S_2;\\
S_1 &=\frac{n}{\ln2}\log 1/\epsilon;\\
S_2 &= \gamma C(n+\epsilon(N-n))=\gamma Cn(1+\epsilon\lambda),\\
\end{cases}
\end{split}
\end{equation*}
which is less than $1.6\cdot S_{\min}$, asymptotically optimal in terms of compression ratio. An illustrative comparison between our algorithm and the theoretical lower bound is depicted in~\autoref{pic:theory}.

% If we set $\epsilon=(\ln^22\gamma C\lambda)^{-1}\leqslant 1$ we can prove the overall space cost
% \begin{equation*}
%     \begin{split}
%             \begin{cases}
%             S   &= S_1 +S_2;\\
%             S_1  &=\frac{n}{\ln2}\log 1/\epsilon;\\
%             S_2  &= \gamma Cn(1+\epsilon\lambda).\\
%             \end{cases} 
%     \end{split}
% \end{equation*}
%  is smaller than $1.6\cdot S_{\min}$. An illustrative comparison between our algorithm and the theoretical lower bound is shown in \bb{Figure \ref{pic:theory}}.

\subsection{Locality Optimization: High Computation Throughput}\label{locality}

While our design described in~\autoref{bloom_filter} enables homomorphic and lossless compression with a high compression ratio, excessive hash computation and random memory access lead to frequent cache misses, limiting the system throughput. 
To address this problem, we propose to perform computations in a batched manner to enhance the locality of memory access, hence reducing cache misses. 
% In~\autoref{peeling_theory}, we have already proved the algorithm exhibits linear computational complexity; Here, we further enhance its speed by locality optimization.

Our solution is to group every $c=1024$ consecutive parameter as a batch, where each batch shares the same index.\footnote{We choose $c$ based on the GPU architecture we use in our experiments. Specifically, the NVIDIA GPUs we use support up to 1024 threads per GPU block.}  
Correspondingly, we reshape the original data $\vec{X}$, the count sketch $\vec{Y}$, and the index $\vec{B}$ into matrices, denoted as $\bb{\textit{X}}$, $\bb{\textit{Y}}$ and $\bb{\textit{B}}$, respectively, each having a width of $c$. To ensure a random distribution of non-zero parameters across different columns of the Count Sketch $\bb{\textit{Y}}$ and the index $\bb{\textit{B}}$, we introduce a random bias within the range $[0..c-1]$ to rotate the order of parameters in one batch when mapping. For example, the 2$^{nd}$, 3$^{rd}$, and 4$^{th}$ parameters of $\bb{\textit{X}}_i$ could be mapped to the 22$^{nd}$, 23$^{rd}$, and 24$^{th}$ parameters of $\bb{\textit{Y}}_{h_1(i)}$, the 122$^{nd}$, 123$^{rd}$, and 124$^{th}$ parameters of $\bb{\textit{Y}}_{h_2(i)}$, and the 1022$^{nd}$, 1023$^{rd}$, and 1$^{st}$ parameters of $\bb{\textit{Y}}_{h_3(i)}$, respectively. This approach guarantees that, as long as the Count Sketch size $||\bb{\textit{Y}}||_0\geqslant \gamma||\bb{\textit{X}}||_0$ where $\gamma=1.23$, full recovery of $\bb{\textit{X}}$ is achievable with a high probability of $1-O(||\bb{\textit{X}}||_0^{-1})$, thereby preserving the lossless property.

\section{Evaluation}\label{sec:evaluation}

We implement and integrate our homomorphic compression algorithm into the NVIDIA Collective Communications Library (NCCL) and the ATP in-network aggregation framework \cite{lao2021atp} in PyTorch~\cite{paszke2019pytorch}. 
We compare our algorithm with two baselines: (1) NCCL AllReduce aggregation that represents worker-level aggregation and (2) ATP's in-network aggregation that represents in-network aggregation without compression. 

We evaluate our algorithm while training the following models of varying sizes and gradient sparsity (\autoref{dataset}): NCF \cite{he2017neural} with the ml-25m \cite{harper2015movielens} dataset,  LSTM \cite{hochreiter1997long} with the GBW \cite{chelba2013one} dataset, VGG19 \cite{simonyan2014very} with the CIFAR-10 \cite{krizhevsky2009learning} dataset, and BERT-base \cite{devlin2018bert} with the SQuAD \cite{rajpurkar2016squad} dataset.
%We evaluate our algorithm while training the following  DNN models since they have varying sizes and gradient sparsity (\autoref{dataset}): NCF~\cite{he2017neural} with the ml-25m \cite{harper2015movielens} dataset,  LSTM \cite{hochreiter1997long} with the GBW \cite{chelba2013one} dataset, VGG19 \cite{simonyan2014very} with the CIFAR-10 \cite{krizhevsky2009learning} dataset, and BERT-base \cite{devlin2018bert} with the SQuAD \cite{rajpurkar2016squad} dataset.
%, the last of which was used for fine-tuning \dk{unclear what this means}. 
%These models were chosen for their varying sizes and levels of gradient sparsity (\autoref{dataset}), to provide a comprehensive evaluation of our algorithm.

We use two separate clusters to evaluate each implementation due to hardware constraints in our testbed. 

(1) \bb{NCCL-based implementation:} The cluster consists of four GPU servers, each equipped with two NVIDIA A100 GPUs with 80 GB memory, interconnected via a 100~Gbps switch. 
GPUs on the same server are interconnected by NVIDIA's NVLink, and routed with a regular switch. 

(2) \bb{ATP-based implementation:} The cluster has an Intel Tofino programmable switch that runs ATP-based in-network aggregation. It consists of three GPU servers, each equipped with an NVIDIA GeForce RTX 2080 GPU with 12~GB memory, interconnected by a 10~Gbps network.

%It's important to note that in the first cluster, the baseline comparison involves using only the NCCL AllReduce aggregation. In contrast, our approach combines data compression with NCCL AllReduce aggregation. Similarly, in the second cluster, the baseline comparison involves using only the ATP in-network aggregation. In contrast, our approach combines data compression with ATP in-network aggregation. This ensures fair comparison across both environments.

%We divide our evaluation into two parts. In micro-benchmark evaluation, we take VGG19 as a case study, showing how recovery accuracy and throughput are influenced by different compressed data sizes. Experiments on all other models show the same trend. In end-to-end evaluation, we fix the compressed data size at 10\% of the original data size and observe the impact on end-to-end training speed for all four models.

\begin{table*}[]
  \caption{DNN models used in our evaluation.}\label{dataset}
	\centering
 % \fontsize{8pt}{9pt}\selectfont
	\begin{tabular}{lccccc}
		\bottomrule
		\textbf{Model}&\textbf{Task}&\textbf{Dataset}&\textbf{Batch Size}&\textbf{Number of Parameters} & \textbf{Average Sparsity}\\ 
            \hline
        \text{NCF}&\text{Recommendation System}&\text{ml-25m}&\text{1024}&29.7 Million&98.9\% \\
            \hline
        \text{LSTM}&\text{Language Modeling}&\text{GBW}&\text{64}&426 Million&94.5\%\\
            \hline
        \text{VGG19}&\text{Image Classification}&\text{CIFAR-10}&\text{128}&140 Million&30.4\% \\
            \hline
        \text{BERT-base}&\text{Question Answering}&\text{SQuAD}&\text{8}&109 Million&20.8\% \\
            \hline
		\toprule
	\end{tabular}
\end{table*}

% \begin{figure}[h!tbp]
%   \centering
%    \includegraphics[width=1\linewidth]{icml2022/distribution.pdf}
% \caption{Distribution of gradient parameters for the first epoch training. The colored regions represent zero values.}
% \label{pic::distribution}
% \end{figure}

\subsection{Microbenchmarks}
\subsubsection{Impact of compression on training accuracy}

To evaluate the losslessness of our algorithm, we measure the recovery accuracy when using our algorithm in training the VGG19 model on a single worker equipped with one GPU (\ie no aggregation across GPUs).  We perform this experiment to demonstrate the recovery performance of our algorithm by decompressing the parameters immediately after compressing them.

We first evaluate the recovery accuracy in each iteration using the following three metrics:

(1) \bb{Average relative error}: It is an average of the absolute values of the relative differences between original and decompressed parameters.

(2) \bb{Recovery rate}: It quantifies the proportion of parameters that are successfully recovered using the peeling theory (\autoref{peeling_theory}). A compression algorithm is lossless if it achieves a 100\% recovery rate, meaning all parameters are successfully recovered. Note that losslessly recovered data may not exactly match the original data. This discrepancy is primarily due to the inherent imprecision of floating-point computation. The lossless property only ensures that if the precision of floating points approaches infinity, the recovered data would perfectly align with the original data. 

(3) \bb{Recovery iterations}: It denotes the number of iterations required to recover all possible parameters. Although not a measure of accuracy, we include it to demonstrate the property that the recovery process typically finishes within $\log\log n + O(1)= 4.7+O(1)$ iterations (\autoref{peeling_theory}).

\begin{figure}
  \centering
   \includegraphics[width=1\linewidth]{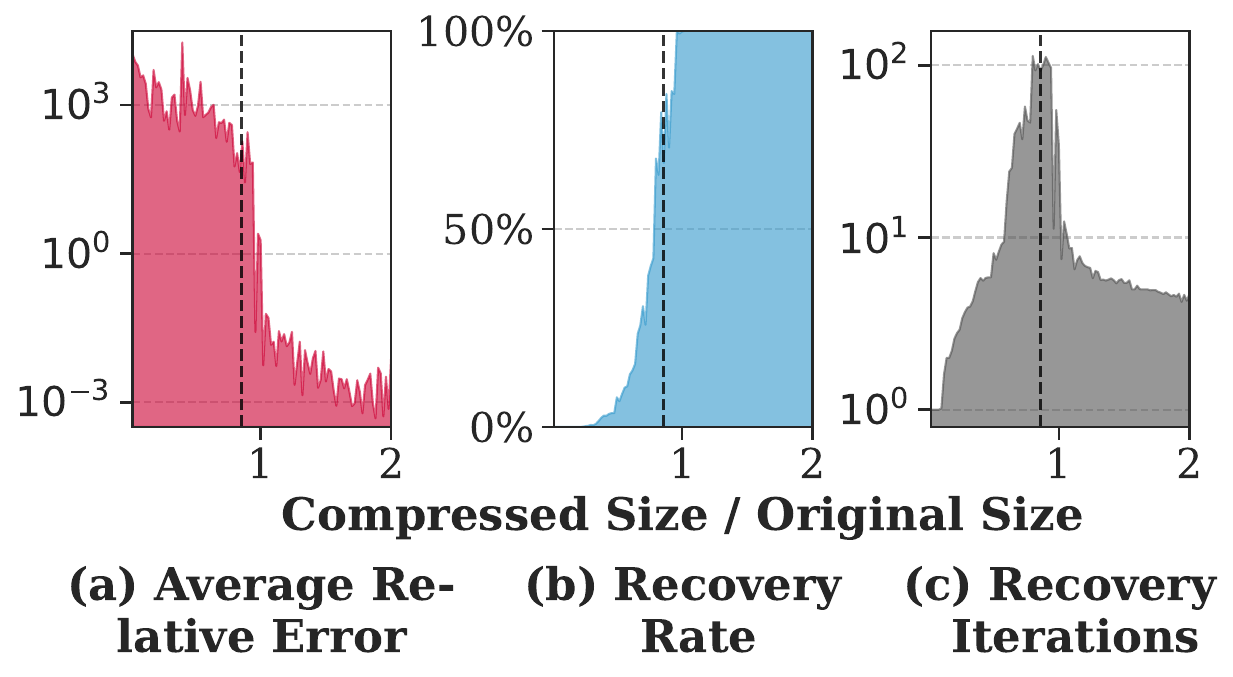}
   \vspace{-0.2in}
\caption{Average relative error, recovery rate, and iterations. }
\label{exp::recovery}
\end{figure}

We vary the compressed data size from 2\% to 200\% of the original data size.\footnote{It may seem odd, but sometimes the compressed data size can be larger than the original data size. This is because when the original data is already dense, compression could actually increase data size by adding redundant information such as indexes in the encoded format.} 
\autoref{exp::recovery} shows the results that once the compressed data size surpasses a certain threshold ($\gamma\times(1-\text{Sparsity})=1.23\times(1-30.4\%)=85.6\%$ of the original data size, see the vertical dotted lines), the average relative error rapidly approaches near zero (\autoref{exp::recovery} (a))  and the recovery rate surges to 100\% (\autoref{exp::recovery} (b)). We also observe a decrease in recovery iterations, indicating a higher throughput of the recovery process (\autoref{exp::recovery} (c)).

\begin{figure}
  \centering
   \includegraphics[width=1\linewidth]{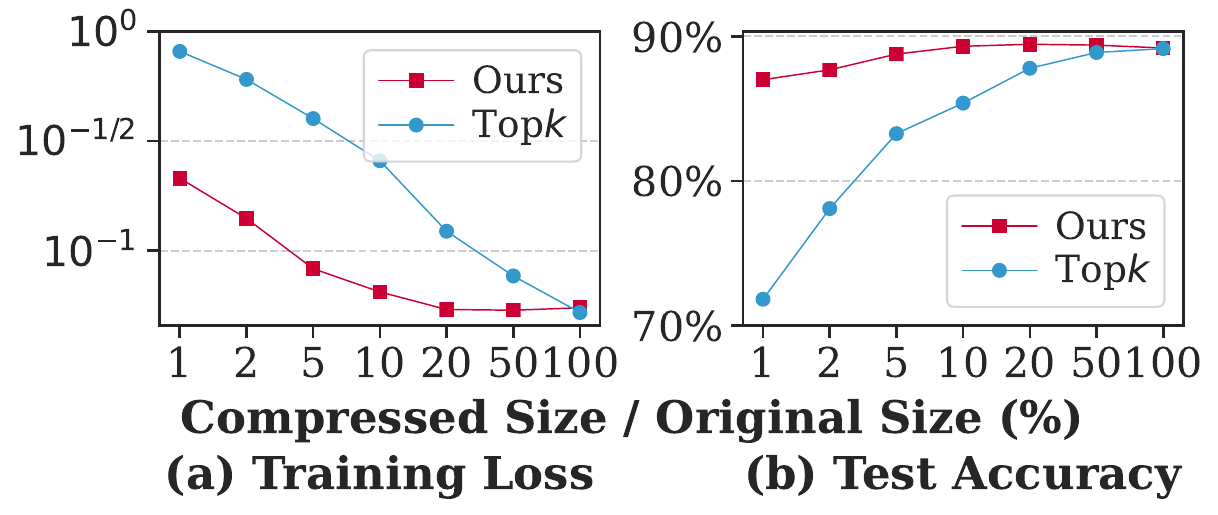}
   \vspace{-0.2in}
\caption{Training loss and test accuracy.}
\label{exp::accuracy}
\end{figure}

Next, we measure the training loss and test accuracy when training the VGG19 model until the model get converged while varying compression ratios.
\autoref{exp::accuracy} shows the results. As the compression ratios decreases (\ie higher recovery rate), training loss decreases while test accuracy increases. This demonstrates why the lossless compression is beneficial even if a lossy approach can preserve convergence.  We also compare our algorithm with the vanilla Top$k$ compression, showing that our algorithm outperforms with the same compression ratio. This is because our algorithm provides an unbiased estimation for the near-zero parameters, while the Top$k$ algorithm simply quantifies them to zero.\footnote{While prior work aims to compensate for the loss of precision by accumulating and transmitting near-zero values every few iterations~\cite{lin2017deep}, it is orthogonal to our solution.}

\subsubsection{Impact of compression on aggregation throughput}

To evaluate the impact of the compression ratio on the aggregation performance, 
we measure the aggregation throughput defined as the volume of aggregated gradients in gigabits per second (\ie Gbps) while varying the compressed data size from 2\% to 100\% of the original data size.\footnote{In our cluster with 8 GPUs, aggregating 8 $\times$ 100 gigabits gradients in one second is counted as 100 Gbps, not 800 Gbps.} Note that the aggregation throughput only takes into account the compression, aggregation, and recovery processes. Other training processes, such as forward and backward propagation, are not included in the aggregation throughput.
In this experiment, we train models using one to four workers, each utilizing either one or two GPUs.

\begin{figure}[]
  \centering
   \includegraphics[width=1\linewidth]{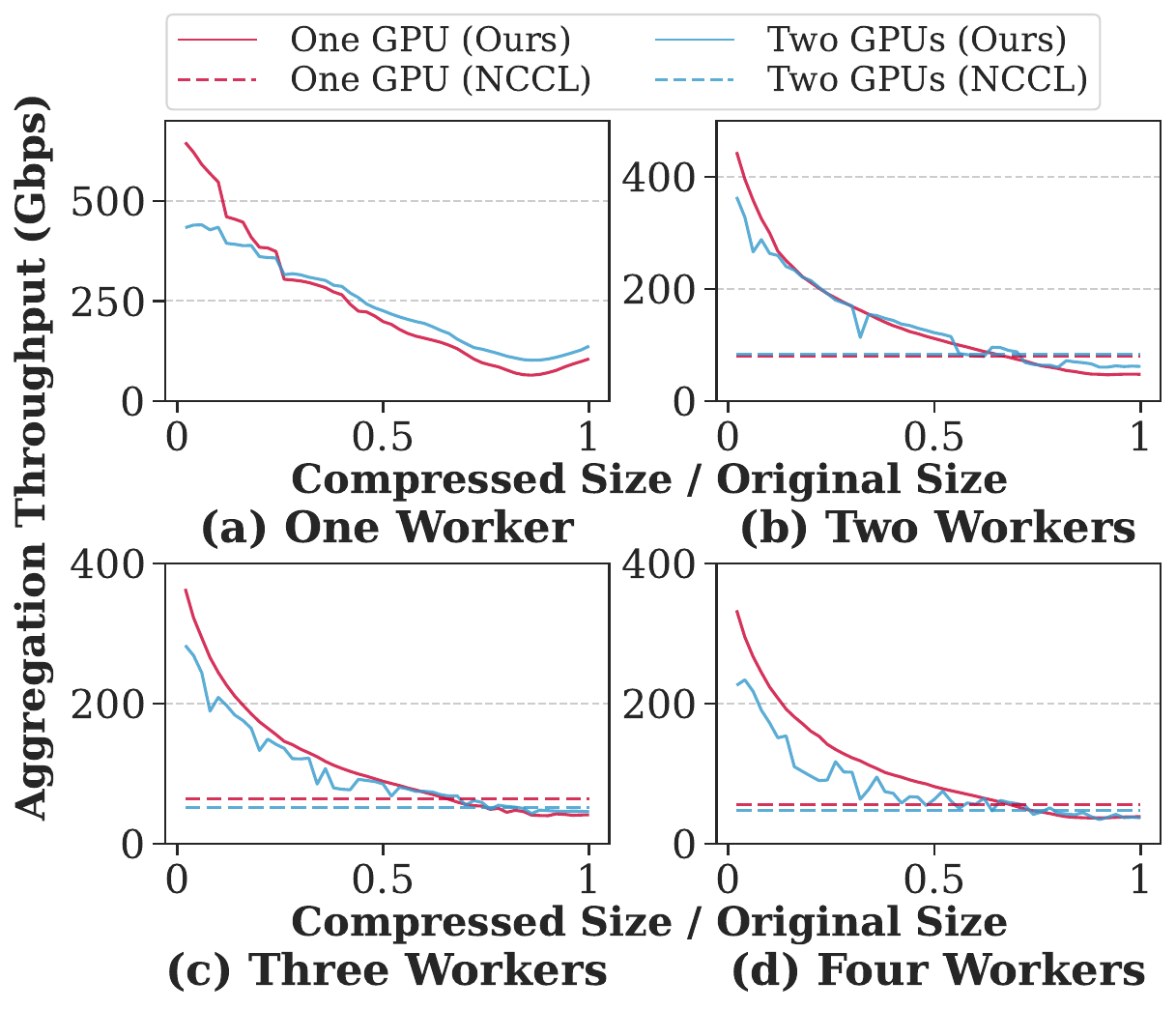}
   \vspace{-0.2in}
\caption{Aggregation throughput of NCCL-based experiments.}
\label{exp::throughput}
\end{figure}

\autoref{exp::throughput} depicts the results. The result from a single worker with a single GPU case shows the throughput of the compression and decompression processes (\autoref{exp::throughput} (a)). At small compressed data sizes, the throughput reaches up to 646~Gbps. 
As the compressed data size increases, the throughput decreases to 64.7~Gbps, mainly due to an increase in memory access and recovery iterations. Interestingly, when the compressed data size reaches 100\% of the original data size (\ie no compression), the throughput increases again. This is likely because of the reduction in recovery iterations.
When using a single worker with two GPUs, if the compressed data size is smaller, the throughput is lower compared to using a single GPU, probably due to the NVLink throughput limit. However, as the compressed data size increases, the dual GPU configuration becomes faster due to the distributed load of the recovery process.

Our algorithm exhibits significant improvements in comparison to the NCCL baseline when multiple workers are employed (\autoref{exp::throughput} (b-d)). When compressed data sizes are small, the algorithm can achieve speedups of up to 4.97$\times$ (4 Workers, 1 GPU) compared to the baseline. However, when data sizes are larger than 60\% of the original data size, the algorithm may experience a reduction in aggregation throughput due to high computational overhead.
%4.28$\times$ (2 Workers, 1 GPU), 4.71$\times$ (3 Workers, 1 GPU), 4.97$\times$ (4 Workers, 1 GPU), 3.36$\times$ (2 Workers, 2 GPUs), 3.43$\times$ (3 Workers, 2 GPUs), and 3.71$\times$ (4 Workers, 2 GPUs) over NCCL baselines. Even at 10\% compressed data size, the algorithm maintains significant speedups of up to $2.73\times$ (2 Workers, 1 GPU), 2.84$\times$ (3 Workers, 1 GPU), 3.01$\times$ (4 Workers, 1 GPU), 2.15$\times$ (2 Workers, 2 GPUs), 3.00$\times$ (3 Workers, 2 GPUs), and 2.59$\times$ (4 Workers, 2 GPUs). However, with larger compressed data sizes exceeding 60\% of the original data size, the algorithm may slow down the aggregation throughput due to high computational overhead.

\begin{figure}
  \centering
   \includegraphics[width=1\linewidth]{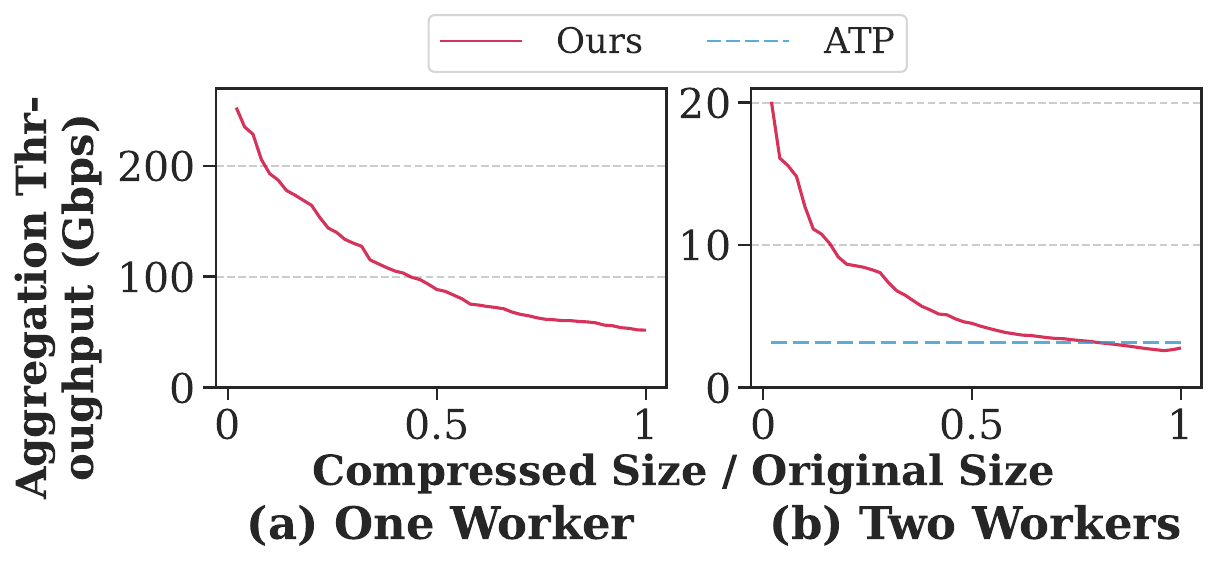}
   \vspace{-0.2in}
\caption{Aggregation throughput of ATP-based experiments.}
\label{exp::throughput_2}
\end{figure}

In ATP-based in-network aggregation experiments, we train models with one or two workers\footnote{Of the three available GPU machines in our testbed, one is used as a parameter server, leaving two available workers.} equipped with a single GPU each (\autoref{exp::throughput_2}). Our findings were similar to those of the NCCL-based experiments, with the maximum specific speedup reaching up to 6.33$\times$.

%We finally focus on speedup and find that on the one hand, when the compressed data size is small, e.g., 2\% of the original data size, our algorithm accelerates up to  $428\%$ (2 Workers, 1 GPU), $471\%$ (3 Workers, 1 GPU), $497\%$ (4 Workers, 1 GPU), $336\%$ (2 Workers, 2 GPUs), $343\%$ (3 Workers, 2 GPUs), and $371\%$ (4 Workers, 2 GPUs) over NCCL baselines. Even if the compressed data size is $10\%$ of the original data size, our algorithm can also accelerate $273\%$ (2 Workers, 1 GPU), $284\%$ (3 Workers, 1 GPU), $301\%$ (4 Workers, 1 GPU), $215\%$ (2 Workers, 2 GPUs), $300\%$ (3 Workers, 2 GPUs), and $259\%$ (4 Workers, 2 GPUs). On the other hand, due to computational overhead, when the compressed data size is large, e.g., exceeding 80\% of the original data size, our algorithm slows down the aggregation process.

\subsection{End-to-end Evaluation}

To evaluate the end-to-end performance of our algorithm, we measure per-iteration and overall speedups. 
Here, we fix the compression data size to 10\% of the original data size. In NCCL-based experiments, we employ four workers, each with two GPUs. In ATP-based experiments, we utilized two workers, each with a single GPU.

\begin{figure}
  \centering
   \includegraphics[width=1\linewidth]{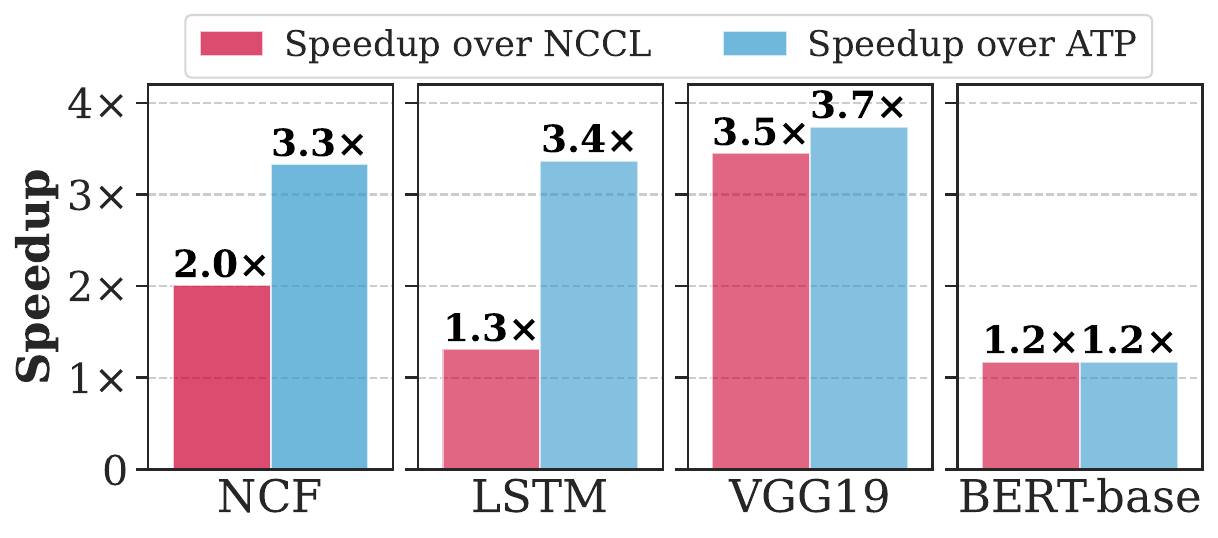}
   \vspace{-0.2in}
\caption{Per iteration speedup. Note that comparing between the red and blue bars would be meaningless because these experiments are conducted on different clusters.}
\label{exp::per_iter_speedup}
\end{figure}

\autoref{exp::per_iter_speedup} shows the per-iteration speedup achieved by our algorithm. We find that on NCF, LSTM, VGG19 and BERT-base models, our algorithm achieves speedup of 2.01$\times$, 1.31$\times$, 3.45$\times$, and 1.17$\times$ over the NCCL baseline, and accelerates 3.33$\times$, 3.37$\times$, 3.74$\times$, and 1.17$\times$ over the ATP baseline. The proportion of computation to communication overhead influences this result. Usually, when communication makes up a larger proportion, our algorithm tends to accelerate more noticeably. It is worth noting that for models such as NCF and LSTM, whose gradient is very sparse, the per-iteration speedup can be approximately the overall speedup. This is because in these cases, our algorithm is verified to recover gradients without loss (\ie with a 100\% recovery rate). However, for VGG19 and BERT-base, the situation is different because our compression is lossy, requiring more training iterations to converge.

\begin{figure}
  \centering
   \includegraphics[width=1\linewidth]{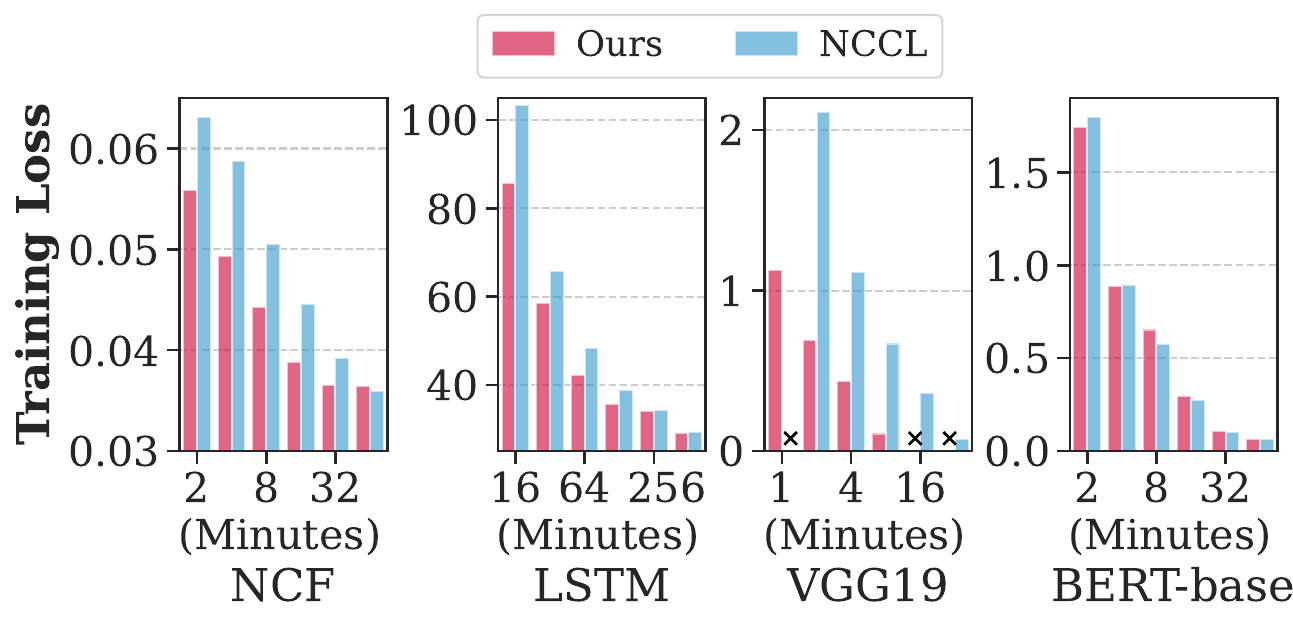}
   \vspace{-0.2in}
\caption{Overall training loss over time.}
\label{overall_loss}
\end{figure}

To evaluate the overall speedups, we measure the training loss.\footnote{The loss functions vary across training tasks. Specifically, we employ perplexity for NCF, PyTorch's \texttt{BCEWithLogitsLoss()} for LSTM, and PyTorch's \texttt{CrossEntropyLoss()} for both VGG19 and BERT-base.}
The experimental results based on NCCL are presented in \autoref{overall_loss}. We compare the training loss achieved by our algorithm and the NCCL baseline over the same amount of training time. On NCF, LSTM and VGG19 models, our algorithm significantly speeds up the training process. Even on BERT-base, which has a non-sparse gradient, our algorithm perform comparably to the NCCL baseline. We observe the same results from ATP-based experiments. 

\section{Conclusion}\label{conclusion}
In this paper, we present a lossless homomorphic algorithm that combines worker-level compression and in-network aggregation to address the communication bottlenecks in DNN training. It uses dual homomorphic structures to create compressed data that preserves all information and leverages algorithms in the area of random graph to recover the original data without losing accuracy. It provides an optimal compression ratio and computational complexity. Our evaluation shows that it improves the aggregation throughput by up to 6.33$\times$ and achieves a 3.74$\times$ speedup in per-iteration training speed in distributed training tasks. We believe that the theories and techniques presented in this paper will inspire more innovative works in the ML community.
% In this paper, we introduce a lossless homomorphic compression algorithm designed for in-network aggregation. This algorithm is built upon minimal assumptions yet achieves nearly optimal compression ratios and computational complexity. We implement our approach using the existing SwitchML and NCCL collective communication API and conduct performance evaluations across three distinct distributed machine learning workloads. Our experimental results demonstrate that our algorithm can accelerate the in-network aggregation process by up to 198\%. Our work amalgamates various traditional network measurement techniques, including sketching, peeling, and membership testing, to address the in-network aggregation challenge. This elegant fusion of techniques is likely to offer fresh insights for future algorithm design.
%%%%%%%%%%%%%%%%%%%%%%%%%%%%%%%%%%%%%%%%%%%%%%%%%%%%%%%%%%%%%%%%%%%%%%%%
% \bibliographystyle{unsrt}
% \bibliographystyle{authoryear}
\bibliography{reference}
\end{document}